# Organometallic Vanadium-Borazine Systems: Efficient One-Dimensional Half-Metallic Spin Filters.

Sairam S. Mallajosyula, Prakash Parida and Swapan K. Pati*

Using density functional theory, we have investigated the electronic and magnetic properties of finite-size as well as infinitely periodic organometallic vanadium-borazine systems ($V_n(B_3N_3H_6)_{n+1}$) for their possible applications in spintronics devices. From our calculations, we find the $V_n(B_3N_3H_6)_{n+1}$ systems to be structurally more stable in comparison to their isoelectronic benzene counterparts ($V_n(C_6H_6)_{n+1}$). All the $V_n(B_3N_3H_6)_{n+1}$ systems are found to be ferromagnetically stabilized, with the infinite one-dimensional $[V(B_3N_3H_6)]_\infty$ wire exhibiting robust half-metallic behaviour. The $V_n(B_3N_3H_6)_{n+1}$ clusters are also found to exhibit efficient spin filter properties when coupled to graphene electrodes.

## Introduction

With the recent noble prize in the year 2007 being awarded for the discovery of the GMR (Giant Magneto Resistance) effect,[1] Spintronics has come forth as an area of active interest, as it holds promises for the next quantum leap in technological advancement.[2] Using the electron spin as a logic bit, this new field has the potential of bringing memory and logic functionalities on the same chip. Half-metals, a new class of compounds which exhibit large carrier spin polarizations, show promising features for spintronics applications, owing to the coexistence of the metallic nature for electrons with one spin orientation and an insulating nature for the other spin orientation.[3] Half metallic behaviour has been reported in three-dimensional materials such as $CrO_2$, manganite perovskites and two-dimensional materials such as graphene nanoribbons.[4] However, with conventional microelectronics moving towards nano-electronics, theoreticians and experimentalists alike are enthusiastically pursuing the search for stable one-dimensional materials which possess half-metallic behaviour.

To this regard, recent experimental studies have predicted ferromagnetic behavior in one-dimensional organometallic Vanadium (V)-Benzene ($C_6H_6$) ($V_n(C_6H_6)_{n+1}$) sandwich clusters.[5] Theoretical calculations predicted a 100% spin-polarized ferromagnetic half metallic ground state for the one-dimensional wire $[V(C_6H_6)]_\infty$, with metallic behavior at the Fermi energy for the minority electrons and an insulating gap for the majority electrons.[6,7] The study also predicted a spin filter effect for a $V_3(C_6H_6)_4$ cluster suspended between a Ni (001) and Co (001) electrodes.[7] Although these gas-phase clusters show interesting properties, it was found that they dissociate on adsorption onto an organic SAM (self-assembled monolayer) matrix, thereby raising question about their stability.[8] Thus, we note that immobilizing these clusters on a metal substrate is still a challenging problem. This turns our attention to finding alternate ways to stabilize the whole electrode-molecule-electrode (EME) system. A way to do this is either to impart stability to the molecular system,[9] or to find an electrode on which the $V_n(C_6H_6)_{n+1}$ clusters can be easily stabilized.

In fact, a recent theoretical study focused on the binding of the $V_n(C_6H_6)_{n+1}$ clusters to Carbon Nanotube electrodes and they found very high transmission spin polarizations, highlighting the efficient spin filter properties of the $V_n(C_6H_6)_{n+1}$ clusters.[10] In search for stable EME systems, we explore the possibility of exploiting the isostrucural similarities between graphene as electrode and $V_n(C_6H_6)_{n+1}/V_n(B_3N_3H_6)_{n+1}$ as molecular systems, where $B_3N_3H_6$ is borazine, an iso-electronic analog of benzene. Such an EME would allow for the stabilization of these Vanadium-molecular clusters on graphene electrodes. This EME system has the added advantage of strong electrode-molecule contact, which is crucial in controlling the device action.

We now focus our attention on the Vanadium (V)-Borazine ($B_3N_3H_6$) ($V_n(B_3N_3H_6)_{n+1}$) sandwich clusters, and their stability in contrast to their Benzene analogs. Borazine, also known as inorganic benzene is a six-$\pi$-electron six-membered ring isoelectronic to benzene.[11] Compared to benzene, the cyclic delocalization of electrons in the borazine ring is reduced due to the large electronegativity difference between boron and nitrogen.[12] The polarity of the B-N bond causes borazine to show a reactivity pattern different from that of benzene. This difference is also highlighted on comparing the stability of the polymorphs of benzene and borazine. While benzene dimers prefer the iso-energetic T-shaped and parallel displaced (PD) conformations over the sandwich structure,[13] borazine on the other hand, prefers the sandwich structure over the T-shaped and PD conformations.[14] This structural stability is due to the favorable charge transfer (CT) interactions. Thus, the point to note is that the formation of linear stacked sandwich structures is preferred in borazine when compared to benzene, which is crucial to the formation of the $V_n(B_3N_3H_6)_{n+1}/V_n(C_6H_6)_{n+1}$ clusters. In fact, from our electronic structure calculations, we predict a robust half-metallic behavior for the infinite 1-D $[V(B_3N_3H_6)]_\infty$ wire. Furthermore, we also show that finite clusters of $V_n(B_3N_3H_6)_{n+1}$ clusters act as nearly perfect molecular spin filters when placed between two graphene electrodes on either side.

## Calculation Method

We have used the SIESTA package for our density functional theory (DFT) calculations.[15] The PBE version of the generalized gradient approximation (GGA) functional is adopted for exchange-correlation.[16] A double-ζ basis set with the polarization orbitals has been included for all the atoms.[17] The semi-core 3p orbital has been included in the valance orbital for the V atom. A real space mesh cut-off of 300 Ry was used for calculating the electronic structure of the $V_n(B_3N_3H_6)_{n+1}$ and $V_n(C_6H_6)_{n+1}$ systems, while a higher mesh cut-off of 400 Ry was used for all the calculations involving graphene layers. Previous studies on one-dimensional systems have shown that GGA results compare well with hybrid functionals such as B3LYP.[6,7,18] All of the calculations were carried out in appropriate super-cells chosen such that the interactions between the neighbouring fragments were negligible. The cell sizes chosen were: $V(B_3N_3H_6)_2$ and $V(C_6H_6)_2$ (15 x 15 x 15 Å$^3$), $V_2(B_3N_3H_6)_3$ and $V_2(C_6H_6)_3$ (15 x 15 x 25 Å$^3$), $V_3(B_3N_3H_6)_4$ and $V_3(C_6H_6)_4$ (15 x 15 x 30 Å$^3$), $V_4(B_3N_3H_6)_5$ and $V_4(C_6H_6)_5$ (15 x 15 x 35 Å$^3$), $[V(B_3N_3H_6)]_\infty$ (20.28 x 20.24 x 7.06 Å$^3$) and $[V(C_6H_6)]_\infty$ (20.04 x 20.02 x 6.92 Å$^3$), Graphene-$V_4(B_3N_3H_6)_3$-Graphene and Graphene-$V_4(C_6H_6)_3$-Graphene (25 x 25 x 30 Å$^3$). Atomic relaxations in all the calculations were performed until the forces on the atoms were not larger than 0.04 eV/Å. To understand the transport properties of the systems under study, we have calculated the transmission function, T(E) at zero bias using the non-equilibrium Green's function methodology extended to spin-polarized systems.[19]

## Results and Discussion

In Fig 1(a), we present the optimized geometry of $V(B_3N_3H_6)_2$. We find that $V(B_3N_3H_6)_2$ adopts the $C_{3v}$ symmetry structure, with the borazine rings being in an eclipsed conformation. It is to be noted that $V(B_3N_3H_6)_2$ can adopt two tautomeric eclipsed structures with the $D_{3h}$ and $C_{3v}$ symmetry, where the difference between the two structures is reflected due to the relative position of the B atoms with respect to the N atoms in the consecutive rings. To analyze the relative stability of the two tautomeric structures, we have performed a conformational space scan, starting form the $D_{3h}$ structure to the $C_{3v}$ structure by rotating one of the borazine ring about the central axis while keeping the other ring fixed, ie, changing the dihedral angle (Φ) between the two borazine rings. The scan was performed in increments of 10°, where the structure of each conformer was relaxed for all the degrees of freedom other than the dihedral angle between the borazine rings, which was fixed. The results of the conformational scan are presented in Fig 1(b). For comparison, we have also presented the results of a similar conformational scan for the $V(C_6H_6)_2$ system in the same graph.

For $V(C_6H_6)_2$, the $D_{6h}$ eclipsed and the $D_{6d}$ staggered structures are found to be nearly iso-energetic, with the eclipsed structure being stabilized over the staggered structure by only 1.01 kcal/mol. While, for $V(B_3N_3H_6)_2$ the two eclipsed structures are found to be inequivalent, with the $C_{3v}$ structure being stabilized over the $D_{3h}$ structure by 8.31 kcal/mol. Also as can be seen from Fig 1(a) for $V(B_3N_3H_6)_2$, the staggered structure, which is the intermediate structure between the two eclipsed structures, is destabilized by 4.55 kcal/mol, in comparison to the $C_{3v}$ structure. Thus, we note that while the benzene rings can rotate freely in $V(C_6H_6)_2$ and thereby can adopt many conformers, only the $C_{3v}$ structure is preferred for $V(B_3N_3H_6)_2$. We find that in the $C_{3v}$ structure, the B atoms are on top of the N atoms of the consecutive rings, thereby favoring charge transfer. While in the $D_{3h}$ structure, the B atoms are on top of the B atoms of the consecutive ring, thereby negating the charge transfer stabilization. Thus, the structural preference is due to favorable charge transfer interactions between the borazine rings in V(B$_3$N$_3$H$_6$)$_2$. It is of interest to note that even in the bulk phase (*h*-BN), the BN layers adopt the *AA'* stacking: the boron atoms in layer *A* being directly above the nitrogen atoms in layer *A'*.[20] The inter-layer separation in h-BN is found to be 3.33 Å, which is comparable to the inter-planar separation of the borazine rings in the $C_{3v}$ (3.50 Å) and $D_{3h}$ (3.45 Å) structures.

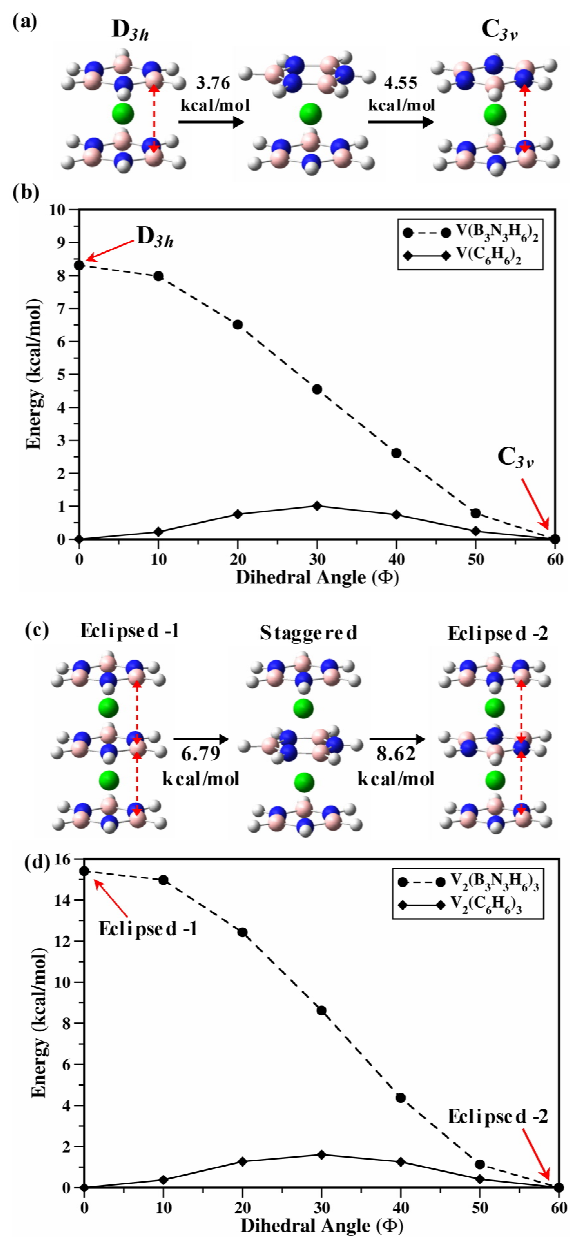

Figure 1. (a) Optimized Geometry of $V(B_3N_3H_6)_2$. Code: dashed red arrows show the arrangement of the B and N atoms. (b) Variation in conformational energy as a function of the dihedral angle for $V(B_3N_3H_6)_2$ and $V(C_6H_6)_2$ (c) Optimized Geometry of $V_2(B_3N_3H_6)_3$; same code as in (a) for arrows. (d) Variation in conformational energy as a function of the dihedral angle for $V_2(B_3N_3H_6)_3$ and $V_2(C_6H_6)_3$. The energy of the eclipsed conformer with the lowest energy is scaled to zero and the Y-axis corresponds to destabilization. Code: dashed line with filled circle: Energy for $V(B_3N_3H_6)_2$ and $V_2(B_3N_3H_6)_3$; straight line with filled circles: Energy for $V(C_6H_6)_2$ and $V_2(C_6H_6)_3$. Energies are reported in kcal/mol and dihedral angles are in degrees.

Interestingly, even in the dimer case (Fig 1(c)) the eclipsed conformer with the B atoms on top of the N atoms in the consecutive rings (Eclipsed-2) is found to be stable over the alternate eclipsed conformer where the B atoms are on top of the B atoms of the consecutive rings (Eclipsed-1). The inter-planar separation between the borazine rings is found to be 3.46 Å and 3.51 Å in the Eclipsed-1 and Eclipsed-2 structures, respectively. Similar to the monomer case, we perform a conformational scan from the Eclipsed-1 to the Eclipsed-2 structure by rotating the middle borazine ring about the central axis while keeping the outer borazine rings fixed. From the results of the conformational scan presented in Fig 1(d), we find that the benzene rings are free to rotate about the central axis in V$_2$(C$_6$H$_6$)$_3$, with the staggered geometry being destabilized over the eclipsed geometry by only 1.6 kcal/mol, ie 0.80 kcal/mol per monomer. On the other hand, the rotation of the borazine ring is highly restricted in V$_2$(B$_3$N$_3$H$_6$)$_3$, with the rotational barriers for the staggered to Eclipsed-2 and Eclipsed-1 to Eclipsed-2 rotation being 8.62 kcal/mol, i.e 4.31 kcal/mol per monomer unit, and 15.41 kcal/mol, i.e 7.70 kcal/mol per monomer unit, respectively. The point to note here is that, the conformational pinning of the monomer and the dimer structures in the eclipsed geometry with the B atoms on top of the N atoms in consecutive rings increases the stability of the V$_n$(B$_3$N$_3$H$_6$)$_{n+1}$ structures and minimizes the chances of the formation of the polymorphic structures, which are common in V$_n$(C$_6$H$_6$)$_{n+1}$ systems.[21]

Before discussing the infinite one-dimensional [V(B$_3$N$_3$H$_6$)]$_\infty$ polymer we present the results of the structural analysis for V$_3$(B$_3$N$_3$H$_6$)$_4$ and V$_4$(B$_3$N$_3$H$_6$)$_5$. The optimized geometries of V$_3$(B$_3$N$_3$H$_6$)$_4$ and V$_4$(B$_3$N$_3$H$_6$)$_5$ are presented in Fig 2(a). As in the monomer and dimer cases, the eclipsed geometry is found to be most stable with the B atoms being on top of the N atoms in the consecutive rings. The inter-planar distance between the borazine rings is found to vary between 3.54 Å and 3.55 Å for V$_3$(B$_3$N$_3$H$_6$)$_4$ and between 3.50 Å and 3.55 Å for V$_4$(B$_3$N$_3$H$_6$)$_5$. Interestingly, we find that the middle borazine ring in V$_4$(B$_3$N$_3$H$_6$)$_5$ is puckered and adopts a chair like conformation which is due to the strong CT interactions with the rings on either side. Such CT interaction leads to the reduction of the aromatic stabilization, which is responsible for the planarity of the borazine ring. Interestingly, such puckering is also noticed for the alkali metal-borazine complexes, where the metal ligand charge transfer causes a loss in aromaticity leading to the puckering of the borazine ring.[22] In Fig. 2(b), we present the optimised geometry of the infinite one-dimensional [V(B$_3$N$_3$H$_6$)]$_\infty$ polymer. We have chosen two V atoms and two B$_3$N$_3$H$_6$ rings per unit cell for our calculations. This is done to analyse the stability of the ferromagnetic and anti-ferromagnetic states. The inter-planar distance between the borazine rings is found to be 3.53 Å which is comparable to the inter-planar separation of 3.46 Å, found between the benzene rings in [V(C$_6$H$_6$)]$_\infty$.

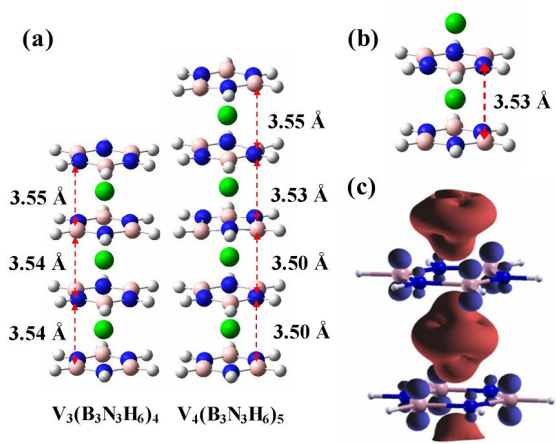

Figure 2. (a) Optimized Geometries of V$_3$(B$_3$N$_3$H$_6$)$_4$ and V$_4$(B$_3$N$_3$H$_6$)$_5$. (b) Optimized Geometry of [V(B$_3$N$_3$H$_6$)]$_\infty$. (c) Spin density map for [V(B$_3$N$_3$H$_6$)]$_\infty$. Code: dashed red arrows show the arrangement of the B and N atoms.

To estimate the stability of the borazine systems, we have calculated the binding energy per monomer for each system. We define the binding energy per monomer (E$_b$) for V$_n$(B$_3$N$_3$H$_6$)$_{n+1}$ as,

$$E_b = [(n+1)E_{Borazine} + nE_V - E_{V_n(Borazine)_{n+1}}] / n$$

where E$_V$ is the energy of the isolated V atom and E$_{Borazine}$ is the energy of the isolated Borazine unit. The E$_b$ for V(B$_3$N$_3$H$_6$)$_2$, V$_2$(B$_3$N$_3$H$_6$)$_3$, V$_3$(B$_3$N$_3$H$_6$)$_4$, and V$_4$(B$_3$N$_3$H$_6$)$_5$ turns out to be 4.22 eV, 3.96 eV, 3.86 eV and 3.86 eV respectively (+ve sign indicating stabilization). Thus, we note that all the borazine clusters are stabilized in comparison to the isolated units. The binding energy for [V(B$_3$N$_3$H$_6$)]$_\infty$ is found to be 3.78 eV. It can thus be seen that with increasing system size, the binding energy does not reduce substantially, the reduction being only 0.44 eV on going from V(B$_3$N$_3$H$_6$)$_2$ to [V(B$_3$N$_3$H$_6$)]$_\infty$. This reduction is less compared to the Vanadium Benzene systems, where the binding energy is found to reduce by 0.62 eV on going from V(C$_6$H$_6$)$_2$ (6.52 eV) to [V(C$_6$H$_6$)]$_\infty$ (5.90 eV).

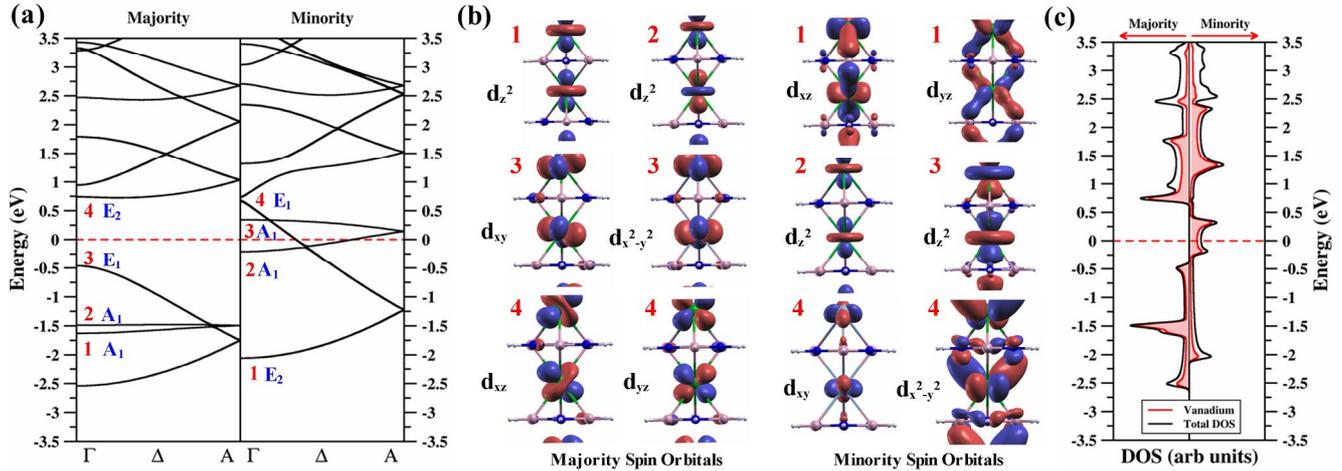

Figure 3. (a) Spin resolved band structure for $[V(B_3N_3H_6)]_\infty$. The plot is scaled for $E_F$ to lie at 0 eV. Important orbitals have been labeled and the corresponding orbitals have been plotted. (b) Orbital plots of $[V(B_3N_3H_6)]_\infty$ calculated at the $\Gamma$ point for the bands labelled in the band structure plot. (c) DOS and pDOS plot for $[V(B_3N_3H_6)]_\infty$. The plot is scaled for $E_F$ to lie at 0 eV. Code: Total DOS (Black line), pDOS from the V orbitals (Red line with shaded area).

Next, we investigate the magnetic interactions present in the borazine systems. We find all the $V_n(B_3N_3H_6)_{n+1}$ (n>1) systems to be ferromagnetically stabilized and in a magnetically polarized state. To estimate the ferromagnetic stability, we calculate the energy difference ($\Delta E$) between the ferromagnetic (FM) and the anti-ferromagnetic (AFM) state. The ferromagnetic stability ($\Delta E$) for $V_2(B_3N_3H_6)_3$, $V_3(B_3N_3H_6)_4$, $V_4(B_3N_3H_6)_5$ and $[V(B_3N_3H_6)]_\infty$ turns out to be –3.6 meV, -9.7 meV, -148.1 meV and –111.6 meV respectively (-ve sign indicates stability of the FM state). As can be seen, the ferromagnetic stability increases with the system size. Interestingly, we find that the total magnetic moment (in Bohr magneton, $\mu_B$) for these systems increases linearly from $V(B_3N_3H_6)_2$ to $V_4(B_3N_3H_6)_5$ in integral multiples of $\mu_B$ with the number of vanadium atoms. Even for the one-dimensional infinite polymer $[V(B_3N_3H_6)]_\infty$, we find an integral magnetic moment of 1 $\mu_B$ per V atom in the unit cell.

To understand the origin of the integral magnetic moment, we present the spin-density map for $[V(B_3N_3H_6)]_\infty$ in Fig 2(c). As can be seen, the V atoms carry large magnetic moments while the $B_3N_3H_6$ molecules are also magnetically polarized to a small extent. In fact, we find a positive magnetic moment of +1.31 $\mu_B$ on the V atoms, while the $B_3N_3H_6$ molecule has a small negative magnetic moment (-0.31 $\mu_B$), which is distributed over the B and N atoms. The B atoms are found to contribute more to the negative moment (-0.084 $\mu_B$ per B) when compared to the N atoms (-0.019 $\mu_B$ per N). This is due to the fact that the lone pair on N is highly localized and does not take part in the electron delocalization. Thus, while for $[V(C_6H_6)]_\infty$ we have a uniform distribution of the magnetic moment on the benzene ring, for $[V(B_3N_3H_6)]_\infty$ we have a non-uniform distribution of the magnetic moment on the borazine ring. However, to understand the magnetic moments in nonmagnetic molecular systems, like $B_3N_3H_6$, we look at the interaction picture in more details.

To identify the orbitals in the composite system which contribute to the finite magnetic moments, we present the spin-polarized band structure for $[V(B_3N_3H_6)]_\infty$ in Fig 3(a). As can be seen, in the periodic composite, the strong crystalline field splits the vanadium 3d levels into a singlet $A_1$ and two doublets, of $E_1$ and $E_2$ symmetry. Due to strong hybridization, the states with the $E_2$ symmetry move far away from the Fermi ($E_F$) level, while the levels with the $A_1$ and $E_1$ symmetry remain closer to the $E_F$. The symmetry of the levels is clearly seen from the orbital plots presented in Fig 3(b). From the orbital plots, we find that the singlet $A_1$ levels have contribution form the $d_{z^2}$ orbitals, while the doublet $E_1$ and $E_2$ levels have contributions from the $d_{x^2-y^2}, d_{xy}$ and $d_{xz}, d_{yz}$ orbitals, respectively. Here it is important to note that, while the isolated V atom has a $3d^3 4s^2$ valence electronic configuration, due to strong hybridization the vanadium 4s levels are shifted above $E_F$. Thus, the five 3d electrons now occupy the non-degenerate $A_1$ and doubly degenerate $E_1$ levels. The majority spin electrons completely fill the two bands with $A_1$ and $E_1$ symmetry, leading to the opening of an insulating band gap of 1.07 eV in the majority spin channel. On the other hand, the remaining two electrons partially occupy the $A_1$ and $E_1$ bands in the minority spin channel. In fact, due to such partial filling, both the bands cross the $E_F$, closing the minority band gap. The important point here is that we have a coexistence of the metallic and insulating nature for electrons in the minority and majority spin channels, respectively, leading to a half-metallic behaviour for the $[V(B_3N_3H_6)]_\infty$ system. To analyse the half-metallic behaviour, we present the DOS and pDOS plots in Fig 3(c), where the pDOS has been projected for V orbitals. Note that, the insulating gap of 1.07 eV for the majority spin channel is comparable to the insulating gap of 1.20 eV found for $[V(C_6H_6)]_\infty$. Another important parameter in determining the robustness of the half-metallic behaviour is the position of the highest occupied molecular orbital (HOMO) in the insulating channel with respect to $E_F$ ($\Delta_{HOMO}$). For $[V(B_3N_3H_6)]_\infty$ $\Delta_{HOMO}$ is found to be 0.50 eV.

Interestingly, for [V(C$_6$H$_6$)]$_\infty$ this gap is found to be 0.30 eV. Thus, we note that the HM behavior is more robust in [V(B$_3$N$_3$H$_6$)]$_\infty$ with $\Delta_{HOMO}$ being greater by 0.20 eV for [V(B$_3$N$_3$H$_6$)]$_\infty$ in comparison to [V(C$_6$H$_6$)]$_\infty$. The opening of this gap is due to the strong hybridization of the CT mediated borazine orbitals with the d-orbitals of V that shifts the hybridized orbitals far from E$_F$. From the band structure plots and the spin density plots, it is clear that the bands with $A_1(d_z^2)$ symmetry are primarily responsible for the integral magnetic moment per V atom in the unit cell.

We now investigate the transport properties of the finite size metal composites, which can be experimentally synthesized, and probe them for possible device applications. The isostructural similarities between graphene and borazine along with the strong adsorption of V atoms on graphene facilitates the stabilization and deposition of these V$_n$(B$_3$N$_3$H$_6$)$_{n+1}$ clusters on graphene electrodes.[10,23] Here we give details of the transport properties of a V$_4$(B$_3$N$_3$H$_6$)$_3$ finite cluster adsorbed on graphene sheet electrodes. The Electrode-Molecule-Electrode system is modeled as Graphene-V(V$_2$(B$_3$N$_3$H$_6$)$_3$)V-Graphene,

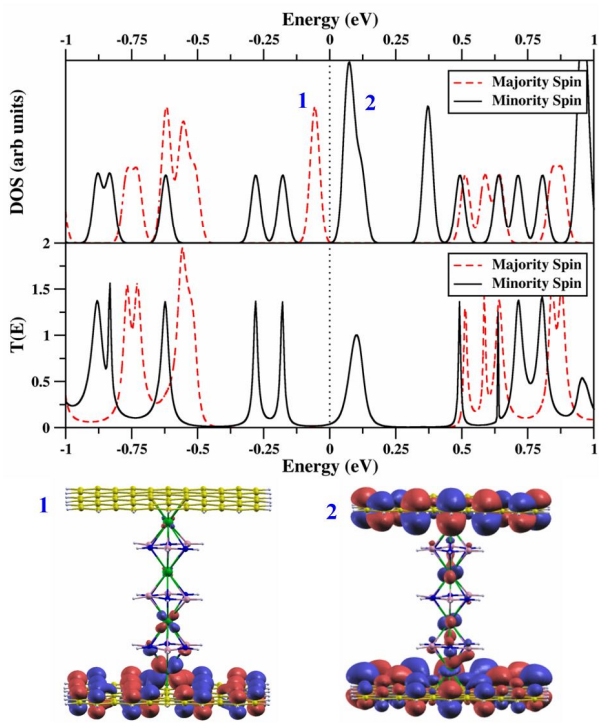

Figure 4. Spin resolved DOS and T(E) plot for Graphene-V(V$_2$(B$_3$N$_3$H$_6$)$_3$)V-Graphene. The plot is scaled for E$_F$ to lie at 0 eV. Important orbitals have been labeled and the corresponding orbitals have been plotted.

We have terminated the cluster by V atoms to improve the binding of the cluster to the graphene electrodes. We have used edge-passivated graphene to model the electrodes, with each electrode consisting of 66 C atoms and 22 H atoms. In Figure 4, we plot the spin polarized DOS and transmission T(E) for the Graphene-V(V$_2$(B$_3$N$_3$H$_6$)$_3$)V-Graphene system. From the DOS plot, we find that the highest occupied molecular orbital (HOMO) is -0.06 eV below E$_F$ in the majority spin channel, while it is -0.17 eV below the E$_F$ in the minority spin channel. On the otherhand, we find that the lowest unoccupied molecular orbital (LUMO) is 0.5 eV above the E$_F$ in the majority spin channel, while it is 0.07 eV above the E$_F$ for the minority spin channel. Note that, the low-energy gaps are nonzero and finite due to the finite sized molecular composites. Since the low-energy molecular orbitals around the Fermi level (E$_F$) govern the transport phenomenon, we restrict our discussion to the HOMO and LUMO of the majority and minority spin channels respectively. As can be seen, there is a strong transmission peak corresponding to the LUMO of the minority spin channel, while the HOMO majority channel does not show any transmission. This can be clearly understood from the analyses of the orbital plots of the HOMO and LUMO of the majority and minority spin channels respectively (see Fig.4). Note that, while the LUMO of the minority spin channel is delocalized over the entire EME system, the HOMO for the majority spin channel is localized on one of the graphene electrode, thereby hindering the transport. For the majority spin channel, we observe strong transmission peaks corresponding to the LUMO (0.5 eV) and the HOMO-1 (-0.5 eV). Since, these orbitals are 0.5 eV away from E$_F$, they do not fall into the bias window and hence do not contribute to transport. Thus, the main point is that only the minority spin electrons take part in the spin transport and hence, at low bias conditions, the finite sized clusters sandwiched between graphene electrodes act as efficient spin filters for advanced spintronics applications.

## Conclusions

To conclude, we have critically examined the electronic and magnetic properties of the Vanadium-Borazine (V$_n$(B$_3$N$_3$H$_6$)$_{n+1}$) clusters. The borazine clusters are found to be structurally stable when compared to their isoelectronic benzene counterparts (V$_n$(C$_6$H$_6$)$_{n+1}$). The structural stability is traced back to the efficient CT interactions present in borazine. All the V$_n$(B$_3$N$_3$H$_6$)$_{n+1}$ clusters are ferromagnetically stabilized. For the infinite [V(B$_3$N$_3$H$_6$)]$_\infty$ polymer system, we predict a stable half-metallic behaviour. Importantly, we find that the half-metallic behaviour is more robust for [V(B$_3$N$_3$H$_6$)]$_\infty$ in comparison to [V(C$_6$H$_6$)]$_\infty$. We also find that finite sized V$_n$(B$_3$N$_3$H$_6$)$_{n+1}$ clusters exhibit efficient spin filter properties when coupled to graphene electrodes. Owing to better structural stability and robust electronic behaviour we conjecture that the V$_n$(B$_3$N$_3$H$_6$)$_{n+1}$ systems would perform as better precursors for advanced spintronics applications.

## Notes and references

*aTheoretical Sciences Unit, Jawaharlal Nehru Center for Advanced Scientific Research, Bangalore 560064, India. Fax: 9180 22082767; Tel: 91 80 22082839; E-mail: pati@jncasr.ac.in*